\documentclass[twocolumn,showpacs,preprintnumbers,amsmath,amssymb,showkeys]{revtex4}

\usepackage{graphicx}
\usepackage{dcolumn}
\usepackage{bm}
\usepackage{bezier}

\begin{document}

\title{Effective action theory of Andreev level spectroscopy}
\author{Artem V. Galaktionov$^{1}$
and Andrei D. Zaikin$^{2,1}$
}
\affiliation{$^1$I.E.Tamm Department of Theoretical Physics, P.N.Lebedev
Physical Institute, 119991 Moscow, Russia\\
$^2$Institute of Nanotechnology, Karlsruhe Institute of Technology (KIT), 76021 Karlsruhe, Germany
}

\begin{abstract}
With the aid of the Keldysh effective action technique we develop a microscopic theory describing
Andreev level spectroscopy experiments in non-tunnel superconducting contacts. We derive an effective impedance of such contacts which accounts for the presence of Andreev levels in the system.
At subgap bias voltages and low temperatures inelastic Cooper pair tunneling is accompanied by
transitions between these levels resulting in a set of sharp current peaks. We evaluate the intensities of such peaks, establish their dependence on the external magnetic flux piercing the structure and estimate thermal broadening of these peaks. We also specifically address the effect of capacitance renormalization in a non-tunnel superconducting contact and its impact on both the positions and heights of the current peaks. At overgap bias voltages the $I-V$ curve is determined
by quasiparticle tunneling and contains current steps related to the presence of discrete Andreev states
in our system.

\end{abstract}


\maketitle

\section{Introduction}
The description of complex systems in terms of the so-called "collective" variables has a long history in condensed matter physics.  An important example of such a variable is the "order parameter field" usually employed for theoretical analysis of phase transitions. A convenience of this approach is guaranteed by the most economic formulation, nevertheless enabling to provide nontrivial results.  Sometimes the correct description can even be constructed phenomenologically, as was the case, e.g., with the celebrated Ginzburg-Landau theory of superconductivity \cite{GL}  justified later on microscopic grounds \cite{Gor}.

Another milestone of this formalism is represented by the Feynman-Vernon influence functional theory \cite{FH} and the related Caldeira-Leggett analysis of quantum dissipation \cite{CL,Weiss}. Within
this description all "unimportant" (bath) degrees of freedom are integrated out and the theory is formulated in terms of the effective action being the functional of the only collective variable of interest.  Both dissipation and superconductivity are combined within the Ambegaokar-Eckern-Sch\"on (AES) effective action approach \cite{AES,SZ} describing macroscopic quantum behavior of metallic tunnel junctions. In this case the collective variable of interest is the Josephson phase, and the whole analysis can be formulated for both superconducting and normal systems embracing various equilibrium and non-equilibrium situations.

Later on it was realized that the AES type-of-approach can be extended to arbitrary (though sufficiently short) coherent conductors, including, e.g., diffusive metallic wires, highly transparent quantum contacts etc. Also in this general case a complete effective action of the system can be
derived both within Matsubara \cite{Z} and Keldysh \cite{SN} techniques, however the resulting expressions turn out to be rather involved
and usually become tractable only if one treats them approximately in certain limits. The character of approximations naturally depends on the problem under consideration. E.g., Coulomb effects on electron transport in short coherent conductors, as well as on shot noise and higher current cumulants can be conveniently studied within the quasiclassical approximation for the phase variable \cite{GZ,GGZ,ns}, renormalization group methods \cite{BN}, instanton technique \cite{N} and for almost reflectionless scatterers \cite{SS,GGZ2}. Some of the above approximations are also helpful for the analysis of frequency dispersion of current cumulants \cite{GGZ2,GGZ3}.

Another type of approximation is realized if one restricts phase fluctuations to be sufficiently small. This approximation may be particularly useful for superconducting contacts with
arbitrary transmissions of their conducting channels. In this case one can derive the effective
action in a tractable form \cite{we} and employ it for the analysis of various phenomena, such as, e.g., equilibrium supercurrent noise, fluctuation-induced capacitance renormalization and Coulomb interaction effects.

An important feature of the effective action \cite{we} is that it fully accounts for the presence of
subgap Andreev bound states in superconducting contacts. In the case of sufficiently short contacts
the corresponding energies of such bound states are $\pm\epsilon_n(\chi )$, where
\begin{equation}
 \epsilon_n(\chi)=\Delta\sqrt{1-T_n\sin^2(\chi/2)},
\label{And}
\end{equation}
$\Delta$ is the superconducting gap, $T_n \leq 1$ defines the transmission of the $n$-th conducting channel and $\chi$ is the superconducting phase jump across the contact. In the tunneling limit $T_n \ll 1$
we have  $\epsilon_n(\chi)\simeq \Delta$ for any value of the phase $\chi$,
i.e. subgap bound states are practically irrelevant in this case. For this reason
such states are missing, e.g., in the AES action \cite{AES,SZ}. On the other hand, at higher
transmission values the energies of Andreev levels (\ref{And}) can be considerably lower than
$\Delta$ and may even tend to zero for fully open channels and $\chi \approx \pi$.
The presence of such subgap states may yield considerable changes in the behavior of
(relatively) transparent superconducting contacts as compared to that of Josephson tunnel junctions.

Recently the authors \cite{Breth1,Breth2} performed experiments aimed at directly detecting Andreev levels by means of microwave spectroscopy of non-tunnel superconducting atomic contacts. In this work we will employ the effective action approach \cite{we} and develop a microscopic theory of Andreev level spectroscopy in superconducting contacts with arbitrary distribution of transmission values $T_n$. As a result of our analysis, we will formulate a number of predictions which would allow for explicit experimental verification of our theory.

The structure of the paper is as follows. In section II we will specify the system under consideration and formulate the problem to be addressed in this work. In section III we will employ our effective action formalism \cite{we} and evaluate the impedance of an effective environment formed by a system involving subgap Andreev levels. These results will then be used in section IV in order to establish the $P(E)$-function for our system and to determine the relative intensity of different current peaks in the
subgap part of the $I-V$ curve. The effect of capacitance renormalization on both the positions and the heights of such peaks will be studied in section V, while in section VI we will address thermal broadening
of these peaks. In section VII we will analyse the $I-V$ curve at larger voltages where
quasiparticle tunneling dominates over that of Cooper pairs. The paper will be concluded in section VIII by a brief summary of our main observations.

 \section{Statement of the problem}

 Following the authors  \cite{Breth1,Breth2} we will consider the circuit depicted in Fig. 1. This circuit can be divided into two parts. The part to the right of the vertical dashed line represents a superconducting loop pierced by an external magnetic flux $\Phi$. This loop includes a  Josephson tunnel junction with normal state resistance $R_N$ and Josephson coupling energy $E_J$ connected to a non-tunnel superconducting contact thereby forming an asymmetric SQUID. The latter contact is characterized by an arbitrary set of transmissions $T_n$ of their transport channels and -- provided
 the superconducting phase difference $\chi$ is imposed -- may conduct the supercurrent \cite{KO}
\begin{eqnarray}
&& I_{S}(\chi )=\frac{e\Delta\sin\chi}{2}\sum_n\frac{T_n}{\sqrt{1-T_n\sin^2(\chi/2)}}
\label{Ichi}
\\ && \times \tanh\frac{\Delta\sqrt{1-T_n\sin^2(\chi/2)}}{2T},
\nonumber
\end{eqnarray}
where $-e$ stands for the electron charge. Below we will assume that temperature $T$ is sufficiently
low $T \ll \Delta$ and we will stick to the limit
\begin{equation}
R_N \ll R_c,
\label{RNRc}
\end{equation}
where $1/R_c=(e^2/\pi )\sum_nT_n$ is the normal state resistance of a non-tunnel contact. In this case the critical current of the Josephson tunnel junction $\propto 1/R_N$ strongly
exceeds that of the non-tunnel superconducting contact $\propto 1/R_c$. In this limit the phase jump across the Josephson junction is close to zero, while
this jump across the non-tunnel contact is
$\chi\approx 2\pi \Phi/\Phi_0$. Here $\Phi_0=\pi c/e$ is the superconducting flux quantum, $c$ is the light velocity and the Planck's constant is set equal to unity $\hbar=1$.

The remaining part of the circuit in Fig. 1 (one to the left of the vertical dashed line) serves as measuring device called a spectrometer \cite{Breth2}. It consists of a voltage biased superconducting tunnel junction with Josephson coupling energy $E_{JS}$ connected to the asymmetric SQUID via a large capacitance $C_0$.

\begin{figure}
\includegraphics[width=9cm]{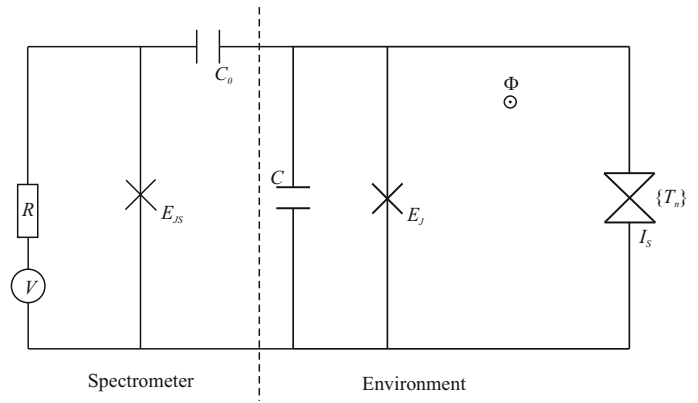}
\caption{The circuit under consideration. The measured system, shown to the right of the dashed line,
represents an asymmetric SQUID comprising a Josephson tunnel junction with resistance $R_N$ and Josephson coupling energy $E_J$ and
a non-tunnel superconducting contact, characterized by an arbitrary set of transmissions $T_n$ of its
conducting channels. The total capacitance $C$ consists of a sum of geometric capacitances of both superconducting junctions $C_\Sigma$ and
also includes the renormalization term from the Josephson element, cf. Eq. (\ref{capren}) below. The superconducting loop is pierced by
the magnetic flux $\Phi$. The measuring device (the spectrometer) is shown to the left of the dashed line.
It incorporates a voltage-biased tunnel junction with Josephson coupling energy $E_{JS}$ connected
to the measured system via a low resistance $R$ and a large capacitance $C_0$.}
\label{steps}
\end{figure}

Assuming that the value $E_{JS}$ is sufficiently small, one can evaluate the inelastic Cooper pair current $I$ across the spectrometer perturbatively in $E_{JS}$.  At subgap values of the applied voltage $V$ one readily finds \cite{Av,Ing}
 \begin{eqnarray}
I=\frac{eE_{JS}^2}{2}\left( P(2eV)-P(-2eV)\right),
\label{pert}
\end{eqnarray}
where
\begin{eqnarray}
P(E)=\int\limits_{-\infty}^\infty dt e^{iEt}\exp\left\{ \frac{4e^2}{\pi}\int\limits_0^\infty \frac{d\omega}{\omega} {\rm Re}\left[ Z(\omega)\right]\right.\nonumber\\
\left.\times
 \left[\coth\frac{\omega}{2T}\left( \cos (\omega t)-1\right)-i\sin(\omega t)\right]\right\}
\label{pe}
\end{eqnarray}
is the function describing energy smearing of a tunneling Cooper pair due to its interaction with
the electromagnetic environment characterized by a frequency-dependent impedance $Z(\omega)$ and temperature $T$. Provided the function $P(E)$ has the form of a delta-function $P(E)\propto\delta(E-E_0)$, the current will be peaked as $I(V)\propto \delta(2eV-E_0)$. This situation is similar to a narrow spectral line on a photoplate, thereby justifying the name of the measuring device.

Coupling of the spectrometer to a single environmental mode (provided, e.g., by an $LC$-contour) was considered in Ref. \onlinecite{Ing}. In this case the environmental impedance takes a simple form
 \begin{equation}
Z_{0}(\omega)=\frac{i\omega}{C((\omega+i0)^2-\omega_0^2)}.\label{plm}
\end{equation}
 Here $C$ is an effective capacitance of the $LC$-contour and $\omega_0$ is the oscillation frequency. As usually, an infinitesimally small imaginary part $i0$ added to $\omega$ in the denominator indicates the retarded nature of the response. Employing Eq. (\ref{pe}) together with the Sokhotsky's formula
 \begin{equation}
 {\rm Im}\,\frac{1}{x+i0}=-\pi\delta(x),
 \end{equation}
in the limit of low temperatures one finds
\begin{equation}
P(E)=2\pi e^{-\rho}\sum_{k=0}^\infty\frac{\rho^k}{k!}\delta(E-k\omega_0),\quad \rho=\frac{4E_C}{\omega_0}.
\label{sm}
\end{equation}
Here and below $E_C=e^2/2C$ is the effective charging energy. Combining Eqs. (\ref{sm}) and (\ref{pert}) we obtain the $I-V$ curve for our device which
consists of narrow current peaks at voltages
\begin{equation}
 2eV=k\omega_0, \quad k=1,2,...
\label{discv}
\end{equation}
The physics behind this result is transparent: A Cooper part with energy $2eV$ that tunnels across the junction releases this energy by exciting the environmental modes. In the case of an environment with a single harmonic quantum mode considered above this process can occur only at discrete set
of voltages (\ref{discv}).

Turning back to the system depicted in Fig. 1, we observe a clear similarity
to the above example of the $LC$-contour. Indeed, the asymmetric SQUID configuration on the right of Fig. 1 plays
the role of an effective inelastic environment for the spectrometer. Bearing in mind the kinetic inductances of both the Josephson element
and the non-tunnel superconducting contact, to a certain approximation this environment can also be viewed as an effective $LC$-contour.
An important difference with the latter, however, is the presence of extra quantum states -- discrete Andreev levels (\ref{And}) --
inside the superconducting contact. Hence, tunneling of a Cooper pair can also be accompanied by upward transitions between these states
and -- along with the current peaks at voltages (\ref{discv}) -- one can now expect the appearance of extra peaks at
\begin{equation}
 2eV=k\omega_0+2\epsilon_n(\chi), \quad k=0,1,2,...
\label{discv2}
\end{equation}

This simple consideration served as a basic principle for the Andreev spectroscopy experiments \cite{Breth1} as well as for their interpretation \cite{Breth2}. While this phenomenological theory \cite{Breth2} correctly captures some important features of the phenomenon, it does not yet allow for the complete understanding of the system behavior, see, e.g., the corresponding discussion in Ref. \onlinecite{Breth2}. Therefore, the task at hand is to microscopically evaluate the function $P(E)$ for the asymmetric SQUID of Fig. 1, which governs the response of the spectrometer to the applied voltage.
In the next section we will describe the effective formalism which will be employed in order to accomplish this goal.

\section{Effective action and effective impedance}

Let us denote the total phase difference across the non-tunnel superconducting contact as $\chi +2\varphi (t)$, where $\chi$ is the constant part determined
by the magnetic flux $\Phi$ and $2\varphi (t)$ is the fluctuating part of the superconducting phase. Assuming that the Josephson coupling energy of a
tunnel junction $E_J$ is sufficiently large one can restrict further
analysis to small phase fluctuations $2\varphi (t) \ll 1$ in both tunnel and non-tunnel contacts forming our asymmetric SQUID.
The total action $S$ describing our system consists of three terms
\begin{equation}
S=S_{Ch}+S_J+S_{sc},
\label{sum}
\end{equation}
describing respectively the charging energy, the Josephson tunnel junction and the non-tunnel superconducting contact. In what follows we will stick to the Keldysh representation of the action in which case it is necessary to consider the phase fluctuation variable on two branches of the Keldysh contour, i.e. to define $\varphi_1(t)$ and $\varphi_2(t)$. At subgap frequencies the sum of the first two terms in Eq. (\ref{sum}) reads
\begin{equation}
S_{Ch}+S_J= -\int dt\varphi_-(t)[\ddot \varphi_+(t)/(2E_C)+4E_{J}\varphi_+(t)].
\label{sum2}
\end{equation}
Here, as usually, we introduced the so-called "classical" and "quantum" phases $\varphi_+(t)=(\varphi_1(t)+\varphi_2(t))/2$, $\varphi_-(t)=\varphi_1(t)-\varphi_2(t)$ and defined
an effective capacitance
\begin{equation}
C=C_{\Sigma}+\frac{\pi}{16 \Delta R_N},
\label{capren}
\end{equation}
which accounts for the renormalization of the geometric capacitance $C_{\Sigma}$ due to fluctuation effects in the Josephson junction \cite{SZ}. The above expansion of the total effective action in powers of (small) phase fluctuations remains applicable for
\begin{equation}
E_{J} \gg E_C.
\end{equation}

Expanding now the action $S_{sc}$ around the phase value $\chi$, we obtain \cite{we}
\begin{equation}
iS_{sc}=-\frac{i}{e}\int\limits_0^t dt'I_S(\chi)\varphi_-(t')+ iS_R-S_I, \label{finalsS}
\end{equation}
where $I_{S}(\chi)$ is defined in Eq. (\ref{Ichi}) and
\begin{eqnarray}
S_R&=&\int\limits_0^t dt'\int\limits_{0}^{t}dt''
{\cal R}(t'-t'') \varphi_-(t')\varphi_+(t''), \label{SRRR}\\
S_I&=&\int\limits_{0}^{t}dt'\int\limits_{0}^{t}dt''
{\cal I}(t'-t'') \varphi_-(t')\varphi_-(t'').
\end{eqnarray}
Both kernels ${\cal R}(t)$ and ${\cal I}(t)$ are real functions related to each other via the fluctuation-dissipation theorem. Defining the Fourier transform of these two kernels respectively as ${\cal R}_\omega={\cal R}'_\omega+i{\cal R}''_\omega$ and ${\cal I}_\omega$ (having only the real part), we obtain
\begin{equation}
{\cal R}''_\omega=2 {\cal I}_\omega \tanh\frac{\omega}{2T}.
\label{FDT}
\end{equation}
The action (\ref{finalsS}) results in the following current through the contact \cite{we}
\begin{equation}
I=I_{S}(\chi)-e \int d t'{\cal R}(t-t')\varphi_+(t')+\delta I(t).
\label{Langevin}
\end{equation}
Here $\delta I(t)$ is the stochastic component of the current. In the non-fluctuating case $\dot\varphi_+(t)=eV(t)$, and Eq. (\ref{Langevin}) defines the current-voltage relation.

The explicit expression for the kernel ${\cal R}(t)$ contains three contributions \cite{we}: One of them originates from the subgap Andreev bound states, another one describes quasiparticle states above the gap and, finally, the third term accounts for the interference between the first two. As here we are merely interested in the subgap response of our system, below we will
specify only the part of the kernel ${\cal R}$ governed by the Andreev bound states.
In the limit of low temperatures it reads (cf. Eqs. (A3), (A5) in Ref. \onlinecite{we}):
\begin{equation}
{\cal R}_\omega=\sum_n\frac{\gamma_n}{4\epsilon_n^2(\chi)-(\omega+i0)^2},
\label{rand}
\end{equation}
where, as before, the summation is taken over the conducting channels of the superconducting contact and
\begin{equation}
\gamma_n=4T_n^2(1-T_n) \frac{\Delta^4}{\epsilon_n(\chi)}\sin^4\frac{\chi}{2}\tanh\frac{\epsilon_n(\chi)}{2T}.\label{randg}
\end{equation}

Now we are in a position to evaluate the current through the spectrometer. In the second order in $E_{JS}$ we obtain
\begin{eqnarray}
&& I(V)=\frac{eE_{JS}^2}{2}\int d t \,{\rm Re}\left( e^{2ieVt} \left\langle e^{2i\varphi_1(t)-2i\varphi_1(0)}+\right.\right.\label{peK}\\&&\left.\left.
e^{2i\varphi_2(t)-2i\varphi_1(0)}- e^{2i\varphi_1(t)-2i\varphi_2(0)}-e^{2i\varphi_2(t)-2i\varphi_2(0)}\right\rangle \right),\nonumber
\end{eqnarray}
where the angular brackets imply averaging performed with the total Keldysh action (\ref{sum}). Under
the approximations adopted here this average is Gaussian and it can be handled in a straightforward manner. As a result, we again arrive at Eqs. (\ref{pert}), (\ref{pe}), where the inverse
impedance of our effective environment takes the form
\begin{eqnarray}
\frac{1}{Z(\omega)}=\frac{C\left(\omega^2-\omega_0^2 \right)}{i\omega}+\sum_n\frac{e^2\gamma_n}{i\omega\left[ 4\epsilon_n^2(\chi)-\omega^2 \right]}.\label{mr}
\end{eqnarray}
Here and below $\omega_0=\sqrt{8E_JE_C}$ is the Josephson plasma frequency.

Eq. (\ref{mr}) -- combined with Eqs. (\ref{And}), (\ref{randg}) -- is our central result which will be employed below in order
to evaluate the $P(E)$-function and to quantitatively describe the results of Andreev level spectroscopy experiments.

\section{Intensity of spectral lines}
It is obvious from Eqs. (\ref{pert}), (\ref{pe}) that the positions of the current peaks are determined by zeroes of the inverse impedance (\ref{mr}). Our theory allows to establish both the positions and relative
heights of these peaks.

To begin with, let us assume that only one transport channel with transmission $T_n$ in our superconducting contact is important, while all others do not exist or are irrelevant for some reason. In this case
from Eq. (\ref{mr}) we obtain
\begin{widetext}
\begin{eqnarray}
&& {\rm Re}\left[ Z(\omega)\right]=\frac{\pi}{4C}\left\{\left[\delta(\omega-\sqrt{x_1})+\delta(\omega+\sqrt{x_1})\right] \left[ 1+\frac{4\epsilon_n^2(\chi)-\omega_0^2}{\sqrt{(4\epsilon_n^2(\chi)-\omega_0^2)^2 +(4e^2\gamma_n/C)}}\right]\right.+
\nonumber\\
&& \left. \left[\delta(\omega-\sqrt{x_2})+\delta(\omega+\sqrt{x_2})\right] \left[ 1+\frac{\omega_0^2-4\epsilon_n^2(\chi)}{\sqrt{(4\epsilon_n^2(\chi)-\omega_0^2)^2 +(4e^2\gamma_n/C)}}\right]
\right\},\label{f1}
\end{eqnarray}
\end{widetext}
where
\begin{equation}
x_{1,2}=\frac{4\epsilon_n^2(\chi)+\omega_0^2\mp\sqrt{(4\epsilon_n^2(\chi)-\omega_0^2)^2 +4e^2\gamma_n/C}}{2}.\label{f2}
\end{equation}
These equations demonstrate that close to the "level intersection" point $\omega_0\approx 2\epsilon_n$ an effective
"level repulsion" is controlled by the factor $\gamma_n$ (\ref{randg}). Outside of an immediate vicinity of this point
one can make use of the condition
\begin{equation}
\gamma_n \ll E_C{\rm max}(\omega_0^2,\epsilon_n^2(\chi))
\end{equation}
(which is typically well satisfied for the parameters under consideration) and expand the
square roots in Eqs. (\ref{f1}), (\ref{f2}) in powers of $\gamma_n$. As a result, one finds
\begin{eqnarray}
{\rm Re}\left[ Z(\omega)\right] =\frac{\pi}{2C}\left[ \delta(\omega-\omega_0)+\delta(\omega+\omega_0)
\right.\nonumber\\
\left.+ \frac{2E_C\gamma_n}{\left(\omega_0^2-4\epsilon_n^2 \right)^2}\left( \delta(\omega-2\epsilon_n)+\delta(\omega+2\epsilon_n)\right)\right].\label{rez}
\end{eqnarray}
Introducing the dimensionless expressions
\begin{equation}
\kappa_n=\frac{E_C\omega_0\gamma_n}{\epsilon_n \left( \omega_0^2-4\epsilon_n^2\right)^2},\label{kap}
\end{equation}
we get up to the first order in $\kappa_n$:
\begin{eqnarray}
P(E)=2\pi e^{-\rho(1+\kappa_n)}\sum_{k=0}^\infty\frac{\rho^k}{k!}\left[\delta(E-k\omega_0)\right.\nonumber\\
\left. +\kappa_n \rho \delta(E-k\omega_0-2\epsilon_n)\right].\label{pef}
\end{eqnarray}
Substituting this result into Eq. (\ref{pert}) we recover the $I-V$-curve of our device at subgap voltages which
fully determines the heights of all current peaks.

For instance, Eq. (\ref{pef}) yields the following ratio for the intensities of the two principal (voltage-integrated)
current peaks occurring at the points $2eV=2\epsilon_n$ and $2eV=\omega_0$:
\begin{equation}
\frac{\int\limits_{eV\approx \epsilon_n} I(V) dV}{\int\limits_{eV\approx \omega_0/2} I(V) dV}=\kappa_n\propto\frac{\sin^4\frac{\chi}{2}}{\epsilon_n^2(\chi)\left( \omega_0^2-4\epsilon_n^2(\chi)\right)^2}.
\label{mch}
\end{equation}
This formula determines relative intensities of the spectral lines as a function of
the phase $\chi$ (or, equivalently, the applied magnetic flux $\Phi$) and constitutes a specific prediction of our theory that can be directly verified in experiments. Eq. (\ref{mch}) holds irrespective of the fact that in any realistic experiment the $\delta$-function current peaks can be somewhat broadened by inelastic effects and it applies not too close to the point $\omega_0=2\epsilon_n$. This ratio of intensities is graphically illustrated in Figs. 2 and 3. The parameters of the figures are chosen in such a way, that $\omega_0=2\epsilon_n$ at $\chi\approx \pi/2$. Fig. 3 is characterized by the smaller value of $\gamma_n$. The approximate expression (\ref{mch}) provides a good description away from $\chi\approx\pi/2$ for both figures. It becomes better in the Fig.3, since it corresponds to smaller $\gamma_n$.

\begin{figure}
\includegraphics[width=8.cm]{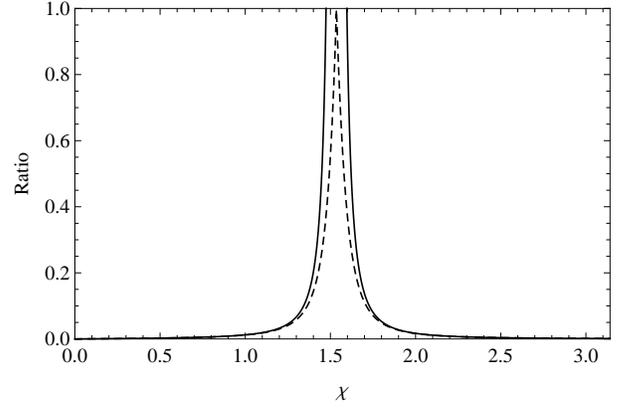}
\caption{The ratio of the intensities of the current peaks at $2eV=2\epsilon_n$ and $2eV=\omega_0$. The parameters are $T_n=0.9$, $\omega_0=1.48\Delta$, $e^2/C=0.4\Delta$. The dashed line results from the exact expression (\ref{f1}), the solid line represents the approximate expression (\ref{mch}). }
\end{figure}

\begin{figure}
\includegraphics[width=8.cm]{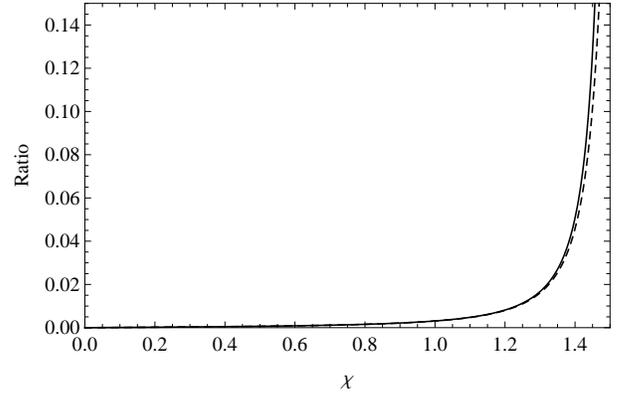}
\caption{The same as in Fig. 2 for $e^2/C=0.1\Delta$.}
\end{figure}

The above consideration can be generalized to the case of several conducting channels in a straightforward manner. For the sake of definiteness let us consider the contacts containing two transport channels with transmissions $T_n$ and $T_m$. In this case Eq. (\ref{f2})
should be modified accordingly. Outside an immediate vicinity of the point $\omega_0=2\epsilon_n$ we obtain the change of the root corresponding to the plasma mode
\begin{equation}
x_1=\omega_0^2+\frac{2E_C\gamma_n}{\left( \omega_0^2-4\epsilon_n^2\right)}+\frac{2E_C\gamma_m}{\left( \omega_0^2-4\epsilon_m^2\right)}+...
\end{equation}
where $...$ stands for higher order in $\gamma_{n,m}$ terms. Similarly, for the other root we get
\begin{eqnarray}
x_2=4\epsilon_n^2+\frac{2E_C\gamma_n}{\left( 4\epsilon_n^2-\omega_0^2\right)}-\frac{4E_C^2\gamma_n^2}{ \left( 4\epsilon_n^2-\omega_0^2\right)^3}\nonumber\\
+\frac{E_C^2\gamma_n\gamma_m}{\left( 4\epsilon_n^2-\omega_0^2\right)^2\left( \epsilon_n^2-\epsilon_m^2\right)}+...
\end{eqnarray}
It also follows that the coefficients in front of the $\delta$-functions in Eq. (\ref{rez}) take the same form in the leading order in $\gamma_{n,m}$. Thus, instead of Eq. (\ref{pef}) we now have
\begin{eqnarray}
P(E)=2\pi e^{-\rho(1+\kappa_n+\kappa_m)}\sum_{k=0}^\infty\frac{\rho^k}{k!}\left[\delta(E-k\omega_0)\right.\nonumber\\
\left.+\kappa_n \rho \delta(E-k\omega_0-2\epsilon_n)+\kappa_m \rho \delta(E-k\omega_0-2\epsilon_m)\right].\label{pef2}
\end{eqnarray}
Close to the intersection point between the plasma mode and one of the Andreev modes the picture will still be governed by Eqs. (\ref{f1}), (\ref{f2}).

Thus, Eq. (\ref{pef2}) demonstrates that the two transport channels just yield "additive" contributions to the $P(E)$-function
describing the asymmetric SQUID under consideration. Along the same lines one can also recover the $P(E)$-function for the case of
more than two transport channels available in the contact.

\section{Capacitance renormalization}

In the above analysis we implicitly assumed that the Josephson plasma frequency $\omega_0$ does not depend on $\chi$. In the interesting for us limit (\ref{RNRc}) this assumption is well justified provided
all channel transmission values $T_n$ remain substantially lower than unity. The situation may change, however, if at least one channel is (almost) open $T_n \approx 1$ and, on top of that, the phase
$\chi$ controlled by the magnetic flux $\Phi$ is driven sufficiently close to $\pi$. In that case capacitance renormalization effects due to phase fluctuations in the superconducting contact may yield
an important contribution which needs to be properly accounted for.

In order to do so we make use of the results \cite{we} where the capacitance renormalization in a superconducting contact with arbitrary distribution of transmissions $T_n$ was investigated in details.
Accordingly, Eq. (\ref{capren}) should in general be replaced by
\begin{equation}
C(\chi )=C_{\Sigma}+\frac{\pi}{16 \Delta R_N}+\delta C(\chi),
\label{capren2}
\end{equation}
where \cite{we}
\begin{eqnarray}
&& \delta C(\chi)=\frac{e^2}{4\Delta}\sum_n \bigg\{ \frac{ 2-(2-T_n)\sin^2(\chi/2)}{T_n\sin^4(\chi/2)} \label{capren3}
\\&&-\left(1-T_n\sin^2(\chi/2) \right)^{-5/2}\bigg[2T_n(T_n-2)\sin^2(\chi/2)
 \nonumber\\&&+5+T_n
+\frac{2-2(1+2T_n)\sin^2(\chi/2)}{T_n\sin^4(\chi/2)}\bigg]\bigg\}.\nonumber
\end{eqnarray}
For any transmission distribution and small phase values $\chi \ll 1$ Eq. (\ref{capren3})
yields
\begin{equation}
\delta C\simeq \frac{\pi}{16 \Delta R_c},
\label{smallchi}
\end{equation}
while for small $T_n \ll 1$ and any $\chi$ one finds
\begin{equation}
\delta C(\chi)=\frac{3\pi}{32 \Delta R_c}\left(1-\frac{\cos\chi}{3} \right).
\end{equation}
In both cases under the condition (\ref{RNRc}) an extra capacitance term $\delta C(\chi)$ in Eq. (\ref{capren2}) can be safely neglected and the latter reduces back to Eq. (\ref{capren}). On the other hand, in the presence
of highly transparent channels with $T_n\approx 1$ Eq. (\ref{capren3}) results in a sharp peak of $\delta C$ at $\chi \to \pi$:
\begin{equation}
\delta C\simeq\frac{e^2}{4\Delta}\sum_n\frac{1}{(1-T_n)^{3/2}}, \label{fd}
\end{equation}
which, depending on the parameters, may even dominate the effective capacitance $C$ at such values of $\chi$. As a result, the plasma frequency $\omega_0$ acquires the dependence on $\chi$ which may become
quite significant for phase values approaching $\chi \approx \pi$. In this case in the results derived
in the previous section one should replace $\omega_0 \to \omega_0(\chi )=\sqrt{8E_{J}E_C(\chi )}$, where
$E_C(\chi )=e^2/2(C+\delta C(\chi))$.

The dependence $\delta C(\chi)$ for various transmission distributions was studied in Ref. \onlinecite{we}
(cf., e.g., Fig. 3 in that paper). One of the important special cases is that of diffusive barriers. In this case the distribution of channel transmissions $T_n$ approaches the universal bimodal form with some channels being almost fully open and, hence, the capacitance renormalization effect should play a prominent role at $\chi \approx \pi$. At such values of $\chi$ one finds \cite{we} $\delta C(\chi)\simeq [\Delta R_c (\pi -\chi )^2]^{-1}$.

It should be emphasized that this capacitance renormalization influences not only Andreev peaks at $eV=2\epsilon_n$, but also the peaks occuring at voltages (\ref{discv}). Namely, as the phase $\chi$ approaches $\pi$ the positions of these peaks are shifted towards smaller voltages (since $\omega_0 \propto 1/\sqrt{C(\chi )}$) while the magnitudes of these peaks decrease (since $\rho \propto 1/\sqrt{C(\chi )}$). Likewise, the magnitudes of principal Andreev peaks $\propto \rho \kappa_n$ may decrease significantly for
$\chi \to \pi$.

\section{Spectral lines width}

Within the framework of our model the width of current  peaks should tend to zero at $T=0$. However, at any nonzero $T$ these peaks become effectively
broadened due to inelastic effects. The corresponding linewidth can be estimated as $\delta \sim 1/(\tilde RC)$, where $\tilde R(T)$ is the effective
resistance of our system which tends to infinity at $T \to 0$ but remains finite at nonzero temperatures. The value $\tilde R(T)$ is controlled by the imaginary part of the kernel ${\cal R}$. It is necessary to include two contributions to this kernel -- one from the non-tunnel superconducting contact (already discussed above) and another one from the Josephson tunnel junction. Accordingly, for the imaginary part of the Fourier component for the total kernel $\tilde {\cal R}$ we have
\begin{equation}
\tilde {\cal R}_\omega''={\cal R}_\omega''+{\cal R}_{J\omega}'',
\end{equation}
where (for $0<\omega<2\Delta$)
\begin{eqnarray}
{\cal R}_{J\omega}''=\frac{1}{e^2 R_N}\int\limits_\Delta^\infty \frac{d\epsilon\left[ \epsilon(\epsilon+\omega)+\Delta^2\right]}{\sqrt{\epsilon^2-\Delta^2}\sqrt{(\epsilon+\omega)^2-\Delta^2}}
\nonumber\\
\times\left[\tanh\frac{\epsilon+\omega}{2T}-\tanh\frac{\epsilon}{2T}\right]\label{imt}
\end{eqnarray}
and ${\cal R}_\omega''$ is obtained from Eq. (\ref{FDT}) combined with Eq. (A1) from Ref. \onlinecite{we}.
As a result, for the subgap region we get
\begin{widetext}
\begin{eqnarray}
&& {\cal R}_\omega''=\sum_n\left\{T_n^{3/2}\left[\tanh\frac{\omega+\epsilon_n(\chi)}{2T}- \tanh\frac{\epsilon_n(\chi)}{2T} \right]\theta(\omega-\Delta+\epsilon_n(\chi))\left| \sin\frac{\chi}{2}\right|\right.\label{ct}\\&& \times\frac{\Delta\left( \omega\epsilon_n(\chi)+\Delta^2(1+\cos\chi)\right)}{2\epsilon_n(\chi)\left((\omega+\epsilon_n(\chi))^2-
\epsilon^2_n(\chi)\right)}\sqrt{(\omega+\epsilon_n(\chi))^2-\Delta^2} \nonumber
\\&&+\frac{T_n}{\pi}\int\limits_\Delta^\infty d\epsilon\frac{\sqrt{\epsilon^2-\Delta^2}\sqrt{(\epsilon+\omega)^2-\Delta^2}}{
(\epsilon^2-\epsilon^2_n)((\epsilon+\omega)^2-\epsilon^2_n)}\left( \epsilon(\epsilon+\omega)+\Delta^2\cos\chi+T_n\Delta^2\sin^2\frac{\chi}{2}\right) \left.\left( \tanh\frac{\epsilon+\omega}{2T}-\tanh\frac{\epsilon}{2T}\right)\right\}.\nonumber
\end{eqnarray}
\end{widetext}
Note that in the lowest order in $T_n$ this expression naturally reduces to that in Eq. (\ref{imt})
(with $R_N \to R_c$). On the other hand, for higher transmission values the difference between the two contributions (\ref{imt}) and (\ref{ct}) become essential: While the former yields the standard thermal factor $\sim \exp(-\Delta/T)$, the latter turns out to be proportional to $\sum_n\exp(-\epsilon_n/T)$ (as long as $\omega+\epsilon_{n}>\Delta$).

It follows from the above consideration that the width of the plasma mode peak can be estimated as
 \begin{equation}
\delta \sim \frac{2E_C \tilde {\cal R}''_{\omega_0}}{\omega_0 },
\end{equation}
whereas the width of the current peak corresponding to the $n$-th Andreev level (away from its intersection with the plasma mode) is
\begin{equation}
\delta \sim \frac{2\kappa_nE_C\tilde {\cal R}''_{2\epsilon_n}}{\omega_0 }.\label{cw}
\end{equation}
with $\kappa_n$ defined in Eq. (\ref{kap}). In the vicinity of the intersection point $\omega_0=2\epsilon_n$ it is necessary to replace $\kappa_n$ by a more complicated expression resulting from Eq. (\ref{f1}).

These estimates demonstrate the crossover from the standard thermal broadening factor $\sim \exp(-\Delta/T)$ to a bigger one $\sim \exp(-\epsilon_n(\chi)/T)$ which accounts for the presence of subgap Andreev levels.

Note that our present consideration is sufficient only in the absence of extra sources of dissipation and under the assumption of thermalization. Both additional dissipation and non-equilibrium effects can further broaden the current peaks beyond the above estimates. Non-equilibrium effects can be captured, e.g., within the effective action formalism
\cite{Kos} which -- being equivalent to that of Ref. \onlinecite{we} in equilibrium -- also allows for non-equilibrium population of Andreev bound states. The corresponding analysis, however, is beyond the frames of the present paper.


\section{Quasiparticle current}
To complete our analysis let us briefly discuss the system behavior at higher voltages $eV >2\Delta$. In this case the $I-V$ curve of our device is determined by quasiparticle tunneling. In the presence
of an inelastic environment one has  \cite{Fal}
\begin{equation}
I_{{\rm qp}}(V)=\int\limits_{-\infty}^\infty \frac{d\omega}{2\pi}\frac{1-e^{- eV/T}}{1-e^{-\omega/T}}P(eV-\omega)I_{{\rm qp}}^{(0)}\left(\frac{\omega}{e} \right).\label{fbs}
\end{equation}
Here $I_{{\rm qp}}(V)$ and $I_{{\rm qp}}^{(0)}$ represent the non-oscillating part of the voltage-dependent quasiparticle current respectively in the presence and in the absence of the environment. At $T \to 0$ the latter is defined by the well-known expression
\begin{eqnarray}
&&I_{{\rm qp}}^{(0)}(V)=\\&& \frac{\Delta}{eR_{NS}}\theta(v-1)\left[ 2vE\left(1-v^{-2}\right)-\frac{1}{v}K\left(1-v^{-2}\right)\right], \nonumber
\end{eqnarray}
where $R_{NS}$ is the normal resistance of the spectrometer junction, $v=eV/2\Delta$ and $E(k),\, K(k)$ are complete elliptic integrals defined as
\begin{equation}
E(k)=\int\limits_0^{\pi/2} d\phi\sqrt{1-k\sin^2\phi}, \; K(k)=\int\limits_0^{\pi/2} \frac{d\phi}{\sqrt{1-k\sin^2\phi}}.
\label{EK}
\end{equation}
\begin{figure}

\includegraphics[width=8.5cm]{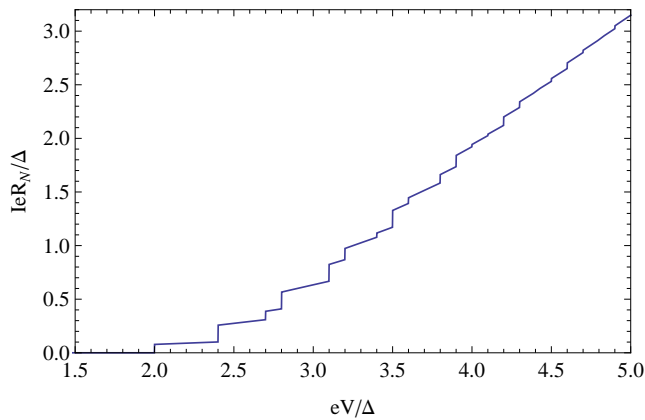}
\caption{Zero-temperature quasiparticle current (\ref{fbs}) with the environment characterized by two quantum modes with frequencies $\omega_1=0.4\Delta$ and $\omega_2=0.7\Delta$. We also set $\rho_1=2$ and $\rho_2=1.$ The current steps are observed at $eV=2\Delta+k\omega_1+l\omega_2$. If $\rho_2$ were much smaller than unity, the steps would be observed at $eV=2\Delta+k\omega_1$ and $eV=2\Delta+k\omega_1+\omega_2$.}
\end{figure}

Combining Eqs. (\ref{fbs})- (\ref{EK}) with the expression for the $P(E)$-function (which is still defined by Eq. (\ref{pef}) with
$\rho \to \rho /4=E_C/\omega_0$) we arrive at the $I-V$ curve which contains two sets of current jumps (steps) at $eV=2\Delta+k\omega_0$ and
$eV=2\Delta+k\omega_0+2\epsilon_n$. This behavior for an effective two mode environment is illustrated in Fig. 4.

\section{Conclusions}

In this work we developed a microscopic theory enabling one to construct a quantitative description of microwave spectroscopy experiments aimed at detecting
subgap Andreev states in non-tunnel superconducting contacts. Employing the effective action analysis \cite{we} we derived an effective impedance of
an asymmetic SQUID structure of Fig. 1 which specifically accounts for the presence of Andreev levels in the system.

At subgap voltages the
$I-V$ curve for the spectrometer is determined by inelastic tunneling of Cooper pairs and has the form of narrow current peaks at voltage values
(\ref{discv}) and (\ref{discv2}). Our theory allows to explicitly evaluate the intensity of these current peaks and establish its dependence on the
external magnetic flux $\Phi$ piercing the system. We also estimated thermal broadening of the current peaks to be determined by the factor
$\sim \exp(-\epsilon_n(\chi)/T)$ rather than by the standard one $\sim \exp(-\Delta/T)$.

In the vicinity of the point $\Phi \approx \Phi_0/2$ and provided at least one of the channel transmissions
$T_n$ is sufficiently close to unity, the positions and heights of the current peaks may be significantly influenced by capacitance renormalization in a
superconducting contact. For instance, the positions of the current peaks can decrease at the flux values $\Phi \approx \Phi_0/2$. We speculate that
this effect could be responsible for experimental observations \cite{Breth1} of such a decrease in one of the samples (sample 3). This sample had about
20 conducting channels some of which could well turn out to be highly transparent, thus providing necessary conditions for substantial $\chi$-dependent
capacitance renormalization.

Finally, we also analyzed the system behavior at overgap voltages $eV >2\Delta$ in which case the $I-V$ curve is mainly determined
by quasiparticle tunneling. The presence of both the plasma mode and Andreev levels results in the sets of current steps
on the $I-V$ curve of our device, as illustrated, e.g., in Fig. 4.

All the above theoretical predictions can be directly verified in future experiments.

\end{document}